\documentclass{sig-alternate-05-2015}

\usepackage{color}
\usepackage{verbatim} 
\usepackage{subfigure} 
\usepackage{multirow}
\usepackage{algorithm}
\usepackage[noend]{algorithmic}
\usepackage{xspace}
\usepackage{graphicx}
\usepackage{epsfig}
\usepackage{amsmath}
\usepackage{amssymb}
\usepackage{times}
\usepackage{float}
\usepackage{enumitem}
\usepackage{balance}
\usepackage[font=bf,belowskip=0pt,aboveskip=5pt]{caption}
\usepackage{lipsum}
\usepackage[colorlinks=false]{hyperref}
\usepackage{url}

\usepackage{epigraph} 

\usepackage{booktabs} 

\usepackage{paralist}

\renewenvironment{enumerate}[1]{\begin{compactenum}#1}{\end{compactenum}}

\newcommand{\hide}[1]{}
\newcommand{\xhdr}[1]{\vspace{1.7mm}\noindent{{\bf #1.}}}

\newcommand{\etc}{\emph{etc.}}
\newcommand{\eg}{\emph{e.g.}}
\newcommand{\ie}{\emph{i.e.}}

\newcommand{\pokemongo}{Pok\'emon Go~}
\newcommand{\pokemongonospace}{Pok\'emon Go}
\newcommand{\pokemon}{Pok\'emon~}
\newcommand{\pokemonnospace}{Pok\'emon}

\graphicspath{{./FIG/}}

\begin{document}

\title{Influence of \pokemongo on Physical Activity: \\Study and Implications}

\numberofauthors{3}
\author{
\alignauthor
Tim Althoff\thanks{Research done during an internship at Microsoft Research.}\\
      \affaddr{Stanford University}\\
      \email{althoff@cs.stanford.edu}
\alignauthor Ryen W. White\\
      \affaddr{Microsoft Research}\\
      \email{ryenw@microsoft.com}
\alignauthor Eric Horvitz\\
      \affaddr{Microsoft Research}\\
      \email{horvitz@microsoft.com}
}

\maketitle

\begin{abstract}

\noindent \textbf{Background: }
Physical activity helps people maintain a healthy weight and reduces the risk for several chronic diseases. Although this knowledge is widely recognized, adults and children in many countries around the world do not get recommended amounts of physical activity. 
While many interventions are found to be ineffective at increasing physical activity or reaching inactive populations, there have been anecdotal reports of increased physical activity due to novel mobile games that embed game play in the physical world. 
The most recent and salient example of such a game is \pokemongonospace, which has reportedly reached tens of millions of users in the US and worldwide.

\noindent \textbf{Objective: }
Quantify the impact of \pokemongo on physical activity. 

\noindent \textbf{Methods: }
We study the effect of \pokemongo on physical activity through a combination of signals from large-scale corpora of wearable sensor data and search engine logs for 32 thousand users over a period of three months. \pokemongo players are identified through search engine queries and activity is measured through accelerometry.

\noindent \textbf{Results: }
We find that \pokemongo leads to significant increases in physical activity over a period of 30 days, with particularly engaged users (\ie, those making multiple search queries for details about game usage) increasing their activity by 1,473 steps a day on average, a more than 25\% increase compared to their prior activity level ($p<10^{-15}$). In the short time span of the study, we estimate that \pokemongo has added a total of 144 billion steps to US physical activity. Furthermore, \pokemongo has been able to increase physical activity across men and women of all ages, weight status, and prior activity levels showing this form of game leads to increases in physical activity with significant implications for public health. In particular, we find that \pokemongo is able to reach low activity populations while all four leading mobile health apps studied in this work largely draw from an already very active population. 
While challenges remain in sustaining engagement of users over the long-term, if
\pokemongo was able to sustain the engagement of its current user base, the game could have a measurable effect on life expectancy, adding an estimated 2.825 million years of additional lifetime to its US users alone.

\noindent \textbf{Conclusions} 
Mobile apps combining game play with physical activity lead to substantial activity increases, and in contrast to many existing interventions and mobile health apps, have the potential to reach activity-poor populations.

\end{abstract}





\section{Introduction}
\epigraph{Those who think they have not time for bodily exercise will sooner or later have to find time for illness.}{Edward Stanley, Earl of Derby, 20 December 1873}

Physical activity is critical to human health. People who are physically active tend to live longer, have lower risk for heart disease, stroke, Type 2 diabetes, depression, and some cancers, and are more likely to maintain a healthy weight (\eg,~\cite{miles2007physical,sparling2000promoting,WHO2010parecommendation}).
Recent analyses estimate that physical inactivity contributes to 5.3 million deaths per year worldwide~\cite{lee2012effect} and that it is responsible for a worldwide economic burden of \$67.5 billion through health-care expenditure and productivity losses~\cite{ding2016economic}.  Only 21\% of US adults meet official physical activity guidelines~\cite{cdc2014pafacts,us2008physical} (at least 150 minutes a week of physical activity for adults), and less than 30\% of US high school students get at least 60 minutes of physical activity every day~\cite{cdc2014pafacts}.
Efforts to stimulate physical activity hold opportunity for improving public health. Numerous studies have called for population-wide approaches~\cite{reis2016scaling,sallis1998environmental}.
However, many interventions have been found to be either ineffective~\cite{dobbins2009school,salmon2007promoting}, to reach only populations that were already active~\cite{dishman1985determinants,marshall2004challenges},
or not to be scalable across varying cultural, geographic, social, and economic contexts~\cite{reis2016scaling}.

Recently, there have been anecdotal reports of novel mobile games leading to increased physical activity, most notably for \pokemongonospace\footnote{\url{http://www.pokemongo.com/}}~\cite{AmericanHeartAssociationNews2016} (other examples include Ingress\footnote{\url{https://www.ingress.com/}} and Zombies, Run!\footnote{\url{https://www.zombiesrungame.com/}}).
\pokemongo is a mobile game combining the \pokemon world through augmented reality with the real world requiring players to physically move around. 
\pokemongo was released in the US on July 6, 2016 and was adopted widely around the world (25 million active users in the US~\cite{surveymonkey2016ususers} and 40 million worldwide~\cite{guardian2016worldwideusers}; 500 million downloads worldwide~\cite{TechCrunch2016wordwidedownloads}).
Due to this massive penetration, \pokemongo can be viewed as intervention for physical activity on a societal-scale. However, its effectiveness for stimulating additional walking has yet to be determined.

\xhdr{Present Work}
We study the influence of \pokemongo on physical activity through a combination of wearable sensor data and search engine query logs for 31,793 users over a period of three months.
Within these users, we identify 1,420 \pokemongo users based on their search activity and measure the effect of playing the game on their physical activity (see Section~\ref{sec:dataset_methods}).
We further compare changes in physical activity for \pokemongo users to changes for large control group of US wearable users and to other leading mobile health apps. 
Lastly, we estimate the impact of \pokemongo on public health.

In summary, our main research questions are:
\begin{enumerate}
    \item Is playing \pokemongo associated with increases in physical activity? How large is this effect and how long does it persist?
    \item Is this effect restricted to particular subpopulations or is it effecting people of all prior activity levels, ages, gender, and weight status?
    \item How does \pokemongo compare to leading mobile health apps in terms of its ability to change physical activity?
    \item How has \pokemongo impacted physical activity in the United States and what is its potential impact on public if the game was able to sustain the engagement of its users?
\end{enumerate}


Our study provides guidance on societal-scale interventions represented by the \pokemongo phenomenon and on the possibilities for increasing physical activity that could be achieved with additional engagement. 
We see this study on \pokemongo as a step towards effectively leveraging games for public health purposes.
Mobile games might not be appealing to everyone and therefore should be seen as a complement rather than a replacement for the interventions considered in the rich body of work on physical activity interventions (\eg,~\cite{dobbins2009school,marshall2004challenges,reis2016scaling,sallis1998environmental,salmon2007promoting,sparling2000promoting}).
To the best of our knowledge, this is the first study to combine large-scale wearable and search sensors to retrospectively evaluate physical activity interventions and the first to study the effect of \pokemongonospace.

\section{Methods}
\label{sec:dataset_methods}

We leverage and combine data from search engine queries with physical activity measurements from wearable devices. 
Specifically, we jointly analyze (1) queries to the Bing search engine mentioning ``pokemon''. We use this to identify which users are likely playing \pokemongo (see Section~\ref{subsec:identifying_users}); 
and (2) physical activity as measured through daily number of steps on the Microsoft Band (see Section~\ref{subsec:dataset_physicalactivity}). 
We jointly use this data to measure differences in physical activity before and after each user shows strong evidence of starting to play \pokemongonospace.

The main study population is 31,793 US users of Microsoft products who have agreed to link data from their Microsoft Band wearables and their online activities to understand product usage and improve Microsoft products. 
In Section~\ref{subsec:identifying_users}, we show that 1,420 users can be classified as \pokemongo players with high confidence. 
We compare changes in physical activity in this population to changes in a control group consisting of a random sample of 50,000 US Microsoft Band users.
For all users, we have self-reported age, gender, height and weight, which we will use in Section~\ref{subsec:effect_size_user_demographics} to estimate the effect of \pokemongo on different groups of users.
Section~\ref{subsec:identifying_users} details how we identify \pokemongo users via strong evidence from search logs 
and Section~\ref{subsec:dataset_physicalactivity} explains the accelerometer-based physical activity data.
Section~\ref{subsec:study_demographics} gives details on study population demographics and Section~\ref{subsec:methods_measuring_pa_impact} explains how we measure the impact of \pokemongo on physical activity.

\pagebreak
\subsection{Identifying \pokemongo Users Through Search Queries}
\label{subsec:identifying_users}

We collected all queries of the 31,793 users between July 6, 2016 (US release date of \pokemongonospace) and August 23, 2016 (date of statistical analysis) that mention the term ``pokemon'' (ignoring capitalization). We then manually annotated the 454 most frequent unique queries in terms of whether they are \textit{experiential}~\cite{paparrizos2016screening,white2016early}; 
that is, the user is very likely playing \pokemongonospace, rather than just being interested in it for some other reason such as following up on news reports or general interest in the game.
This was done by an author familiar with the game manually executing each query and judging whether the query and search engine results provided compelling evidence of someone playing the game.
Examples for experiential and non-experiential queries are given in Table~\ref{tab:example_queries}.

Among the 25,446 users who issued any queries during our time of observation, 1,420 or 5.6\% issued an experiential query for \pokemongonospace. 
This number very closely matches the estimated fraction of regular Pokemon users in the US (5.9\% according to \cite{TechCrunch2016PokemonUserStats}) suggesting that our search-engine based method is effectively detecting a large number of \pokemongo users.
We use the time of each user's first experiential query for \pokemongo as a proxy for the time when they started playing \pokemongo and denote this time as $t_0$.

Note that our method of identifying \pokemongo players through experiential queries can potentially overestimate $t_0$ if players perform these queries several days after starting to play the game, but the opposite is less likely due to the nature of experiential queries targeting specific aspects of game play (see Table~\ref{tab:example_queries}).
However, note that any potential overestimates of $t_0$ lead to more conservative estimates of the effect of \pokemongo since potential game-related increases in activity would be counted as activity before $t_0$ (assuming the effect is non-negative).

\begin{table}[t]
\centering
\resizebox{1.0\columnwidth}{!}{%
\begin{tabular}{ll}
  \toprule
  Non-experiential query & Experiential query \\
   \midrule
    pokemon go  &    pokemon go iv calculator  \\
    pokemon go death san francisco    &    pokemon go teams   \\
    pokemon go robberies    &    how to play pokemon go  \\
    couple sues pokemon go   &    pokemon go guide   \\
    baltimore pokemon accident & pokemon go servers \\
    pokemon games & pokemon go bot \\
    bluestacks pokemon go & pokemon go eevee evolution \\
  \bottomrule
 \end{tabular}
 }
 \caption{
 Representative experiential and non-experiential \pokemongo queries~\cite{paparrizos2016screening}. ``iv'' refers to individual values which are attribute points of \pokemon determining their stamina, attach and strength; ``bluestacks'' refers to a method to play \pokemongo on a desktop computer instead of the intended use in the real world; ``eeevee'' is the name of a \pokemonnospace.  See Section~\ref{subsec:identifying_users} for more details on the 454 features used.
 }
 \label{tab:example_queries}
 \end{table}

\begin{table}[t]
\centering
\resizebox{0.9\columnwidth}{!}{%
\begin{tabular}{lrr}
  \toprule
  Minimum number of exp. & \#Users & \#Days with steps data \\
  \pokemongo queries & & \\
   \midrule
  1  &    792  &  36,141 \\
  2  &    417  &  18,804 \\
  3  &    262  &  11,916 \\
  4  &    199  &   9,132 \\
  5  &    143  &   6,633 \\
  6  &    113  &   5,186 \\
  7  &     85  &   3,819 \\
  8  &     70  &   3,131 \\
  9  &     56  &   2,512 \\
 10  &     50  &   2,218 \\
  \bottomrule
 \end{tabular}
 }
 \caption{
Number of \pokemongo users and number of days of steps tracking for these users included in dataset. We count days up to 30 days before and after each user's first experiential query, and only consider users with at least one day tracked before and after their first experiential query.
 }
\label{tab:dataset_stats_nusers_stepsdays}
 \end{table}

\begin{figure}[t]
\centering
\includegraphics[width=.95\columnwidth]{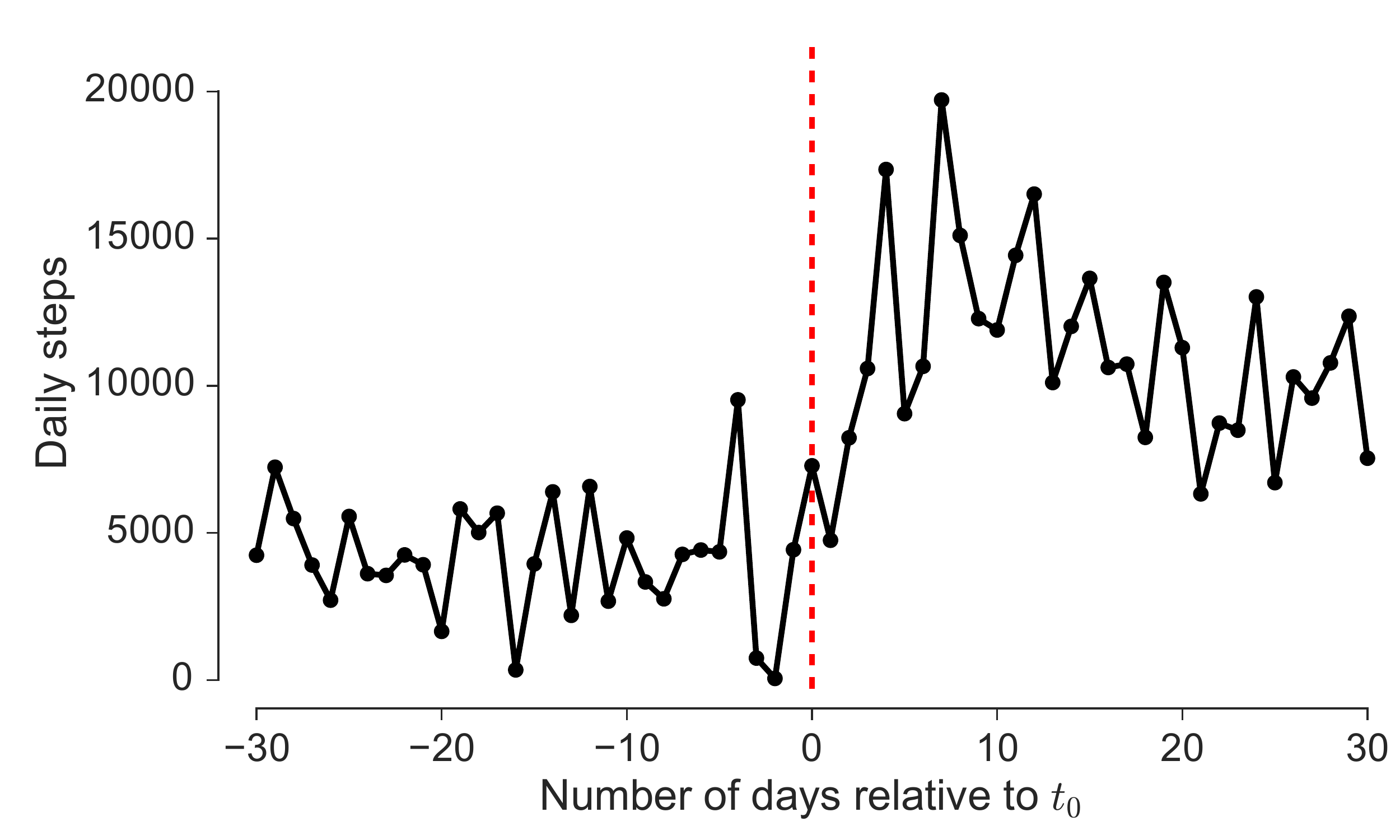}
\includegraphics[width=.95\columnwidth]{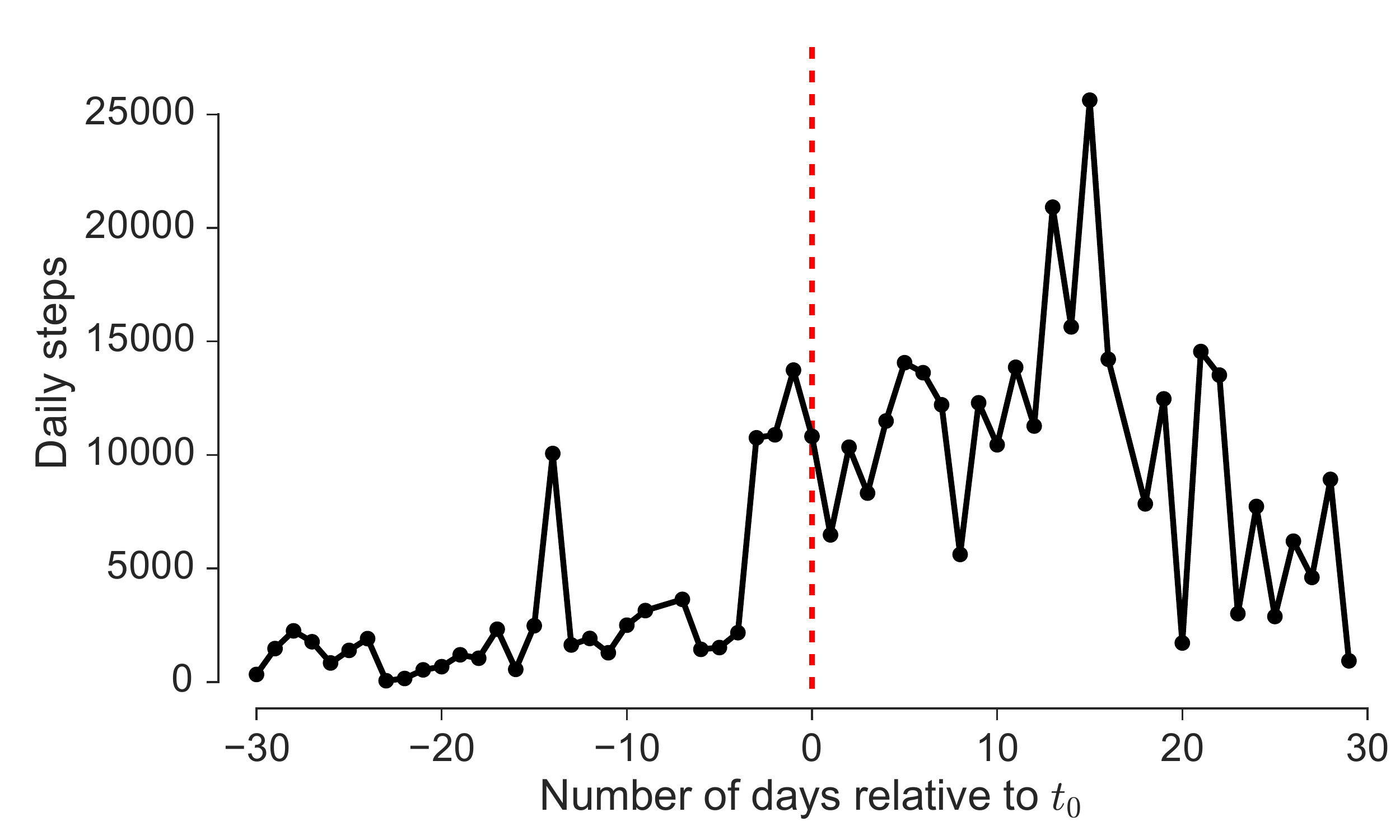}
\caption{Time series of daily steps for two sample users. Both cases show significant increases in daily steps after the first experiential query for \pokemongo ($t_0$). While before $t_0$ both users take less than 5,000 steps a day, after $t_0$ they regularly reach around 15,000 steps a day. 
}
\label{fig:example_user_time_series}
\end{figure}

\begin{table}[t]
\centering
\resizebox{1.0\columnwidth}{!}{%
\begin{tabular}{p{47mm}rr}
  \toprule
  & \pokemongo Users & Wearable Users \\
   \midrule
    \# users & 1,420 & 50,000 \\
    \# users with sufficient activity data & 792 & 26,334 \\
    Median age & 33 & 42\\
    \% female & 3.8 & 25.7\\
    \% underweight (BMI $<$ 18.5) & 1.1 & 1.2\\
    \% normal weight (18.5 $\leq$ BMI $<$ 25) & 34.2 & 31.4\\
    \% overweight (25 $\leq$ BMI $<$ 30) & 36.5 & 38.4\\
    \% obese (30 $\leq$ BMI) & 28.2 & 29.1\\
    Average daily steps overall & 6,258 & 6,435\\
  \bottomrule
 \end{tabular}
 }
 \caption{
    Dataset statistics. 
    Wearable users refers to random sample of US Microsoft Band users.
    We only consider users with at least one day of steps tracking before and after the user's first experiential query.
    BMI refers to body mass index.
 }
 \label{tab:study_population_demographics}
 \end{table}

\subsection{Measuring Physical Activity}
\label{subsec:dataset_physicalactivity}

We seek to measure the change in physical activity before and after the time of the first experiential query for \pokemongonospace, $t_0$, when a user presumably started playing the game (see Section~\ref{sec:effect_on_pa}). 
We measure physical activity through daily steps as recorded by the 3 axis accelerometer/gyrometer of the Microsoft Band.
Accelero\-meter-defined activity measures are preferred over subjective survey-based methods, that have been found to overestimate physical activity by up to 700\%~\cite{tucker2011physical}.
We use steps data from 30 days before the first experiential query ($t_0$) until 30 days after the first experiential query. 
We note that, at the time of this study, very few users had been using \pokemongo for more than 30 days. 
Further note that all \pokemongo users included in our dataset have been using the wearable device for a significant amount of time (median 433 days) such that differences in activity cannot be due to starting to use the wearable device.
Since not every search engine user who we identified as a \pokemongo player is also regularly tracking steps, there are 792 users that tracked steps on at least one day before and after $t_0$ (see Table~\ref{tab:dataset_stats_nusers_stepsdays}). 
Note that the choice of this threshold parameter does not significantly impact our analysis as we find very similar results when restricting our analysis to users tracking for example seven days before and after $t_0$.
We concentrate our analysis on this set of users and compare their activity to the control group described below. 

\xhdr{Control Group}
We further compare the differences in activity in the \pokemongo user population to any changes in the control group, a random sample of US Microsoft wearable users. 
For example, summertime along with improved weather conditions and potential vacation time might be linked to increases in the steps of the control group as well. 
Since there are no experiential queries for any of the control users, we need to define a suitable substitute for $t_0$ for the control group in order to compare both groups.
We will use this reference point $t_0$ to measure changes in physical activity before and after for both the \pokemongo user group as well as control users.
For the \pokemongo users, $t_0$ corresponds to the date of the first experiential query for \pokemongo (\eg, July 6, 2016, or July 7, 2016, \etc).
One could consider using a single point in time $t_0$ for all control users, for example the July 6, 2016 release date of \pokemongonospace. 
However, this choice would temporarily align all control users such that weekend, weather, or other effects could lead to confounding.
In the \pokemongo user group, all users have potentially different $t_0$ based on their first experiential query and therefore such effects are not aligned.
In order, to match observation periods between both groups, we therefore use the exact same distribution of $t_0$ for control users; that is, for each control user, we randomly sample a \pokemongo user and use the same value for $t_0$ for the control user.
This ensures that we will compare physical activity over matching observation periods.

\xhdr{Wear Time}
Furthermore, we also measure the wear time of the activity tracking device for each day in the dataset. Differences in recorded number of steps could potentially stem from simply an increase in wear time rather than an actual increase of physical activity. However, we find that during the study duration the wear time for both \pokemongo and control users was effectively constant with the ratio between the groups changing by less than one percent.
Therefore, we attribute any differences in recorded number of steps to an actual increase in physical activity due to the engagement with \pokemongonospace.

\xhdr{Example Time Series of Physical Activity}
Figure~\ref{fig:example_user_time_series} displays the daily number of steps before and after the user's first experiential query for two example users.
Both users significantly increase their activity after their first experiential query for \pokemongo by several thousand steps each day.
In Section~\ref{sec:effect_on_pa}, we analyze whether this large increase in physical activity is representative of the study population and how it varies across individuals.

\subsection{Study Population Demographics}
\label{subsec:study_demographics}
Demographic statistics on identified \pokemongo users and control users are displayed in Table~\ref{tab:study_population_demographics}.
We find that \pokemongo users are younger than the average user in our wearable dataset, and much less often female.
Furthermore, there is a significant fraction of overweight and obese users, similar to the proportion expected in the US population~\cite{niddk2012obesitystats}.
This fraction of overweight and obese users is very similar in the \pokemongo and control user groups indicating lack of a selection effect based on weight status.
The average activity level of \pokemongo users is below that of the control group indicating that that \pokemongo is attracting users that get less than average activity.
Note that this difference is unlikely to stem from other differences between the two groups since younger users are typically more active than older users and males typically get more physical activity than females~\cite{tudor2004many} (\ie, we would expect a larger number of steps for the \pokemongo group given the other differences).

\subsection{Measuring the Impact of \pokemongonospace}
\label{subsec:methods_measuring_pa_impact}
This section details the methods used to measure the impact of \pokemongo on physical activity.

\subsubsection{Longitudinal Analysis}
\label{subsec:longitudinal_analysis_methods}
We compare the physical activity levels of \pokemongo users to those of the control group population over time in relation to every user's first experiential query ($t_0$). 
Note that we use randomly sampled $t_0$ for users in the control group (see Section~\ref{subsec:dataset_physicalactivity}).

We measure the average number of steps over a period of 30 days before the first experiential query until 30 days after the first experiential query.
Note that on some days a user might not have recorded any steps and we ignore this user on that day.
We measure this average activity separately for the \pokemongo user group and the control group.
To improve graph readability, we smooth the daily average activity through Gaussian-weighted averaging with a window size of seven days, but we report statistical tests on the raw data. 
We estimate 95\% confidence intervals through a bootstrap with 500 resamples~\cite{efron1994introduction}.

\subsubsection{Dose-Response Relationship between \pokemongo and Physical Activity}
\label{subsec:quantifying_effect_size_methods}
Dose-response relationships between the amount of physical activity and various health outcomes have been well established~\cite{dunn2001physical,lee2001physical}.
We expect that high engagement with \pokemongo would be reflected in a larger number of experiential queries.
Particularly engaged users might also exhibit larger increases in physical activity.
We quantify the exact effect sizes for these increases and study this potential dose-response relationship between the \pokemongo related engagement on a search engine and real-world physical activity.
We measure the difference in the average number of daily steps across all users and days for the 30 days before versus 30 days after each user's first experiential query as the effect size.

\subsubsection{Does Everyone Benefit?}
\label{subsec:effect_size_user_demographics_methods}

%
We measure the effect on individual users' physical activity after starting to play \pokemongo and relate the magnitude of this effect to demographic attributes of the user including  age, gender, weight status (body mass index; BMI), and prior activity level.
We investigate whether only certain user groups are benefiting from the game or whether the potential health benefits might apply more widely to the game's user population.
%
We estimate the effect of playing \pokemongo on each individual user defined as the difference in the average number of daily steps 30 days before and 30 days after the first experiential query. 
We include only \pokemongo users with at least seven days of steps tracking before and after this event to reduce noise and apply the same requirement to the control group.
These constraints result in 677 \pokemongo users and 26,334 control users.

\pagebreak
\subsubsection{Comparison to Existing Health Apps}
\label{subsec:app_comparison_methods}

We compare the effect of \pokemongo to the effect of other mobile health apps.
The Microsoft Band can be connected to other fitness and health applications and we have data on when these connections first happen (\ie, explicit knowledge of $t_0$ for users of these apps).
We study four leading mobile health applications with anonymized names for legal reasons.
These apps regularly are rated among the top health apps on both iOS and Android platforms and represent the state-of-the-art in consumer health applications.
%
Again, we measure the number of daily steps 30 days before a user starts using one of these apps until 30 days after.
We only include users that started using the health applications after July 1, 2016 to control for seasonal effects and make the data comparable with our \pokemongo user group.
We only include users that were tracking steps on at least 7 days before and after the first experiential query (for \pokemongo group) or first connecting the health app (for the comparison groups).
For the four apps, 1,155 users are included for app A, 313 for app B, 625 for app C, and 296 users for app D.
Note that these users had been using the wearable device for a significant amount of time before connecting to the health app (median time in days for the four apps are 87, 57, 103, and 76 days, respectively).
Therefore, any differences in average activity are likely due to the connected health app rather than cumulative effects of starting to use a wearable activity tracker.

\subsubsection{Estimating the Public Health Impact of \pokemongonospace}
\label{subsec:public_health_impact_methods}

In order to quantify the effect of \pokemongo on public health, we estimate 
(1) how many steps were added to US users' physical activity during the first 30 days, 
(2) how many users met physical activity guidelines before and after \pokemongonospace,
and (3) the potential impact on life expectancy if \pokemongo could sustain the engagement of its users.

The official physical activity guidelines~\cite{cdc2014pafacts,us2008physical} are equivalent to approximately 8,000 daily steps~\cite{tudor2011stepsguidelines, tudor2011accelerometer}. 
Only 21\% of US adults meet these guidelines.
We use all users tracking steps at least seven days before and after their first experiential query for \pokemongonospace.
We then measure the fraction of users with more than 8,000 average daily steps both 30 days before and after the first experiential query. 
This analysis is repeated for \pokemongo users with at least one and at least ten experiential queries, and the control group.

If there is a substantial impact on physical activity, \pokemongo could have a measurable impact on US life expectancy due to well-established health benefits of physical activity on heart disease, stroke, Type 2 diabetes, depression, some cancers, obesity, and mortality risk~\cite{ding2016economic,lee2012effect,miles2007physical,sparling2000promoting,WHO2010parecommendation}.
If we assume that \pokemongo users would be able to sustain an activity increase of 1,000 daily steps,
this would be associated with a 6\% lower mortality risk. 
Using life-table analysis similar to~\cite{lee2012effect} based on mortality risk estimates from \cite{dwyer2015objectively} and the United States 2013 Period Life Table~\cite{usssa2013periodlifetable} we estimate the impact on life expectancy based on this reduction of mortality risk.

\begin{figure}[!ht]
\centering
\includegraphics[width=0.999\columnwidth]{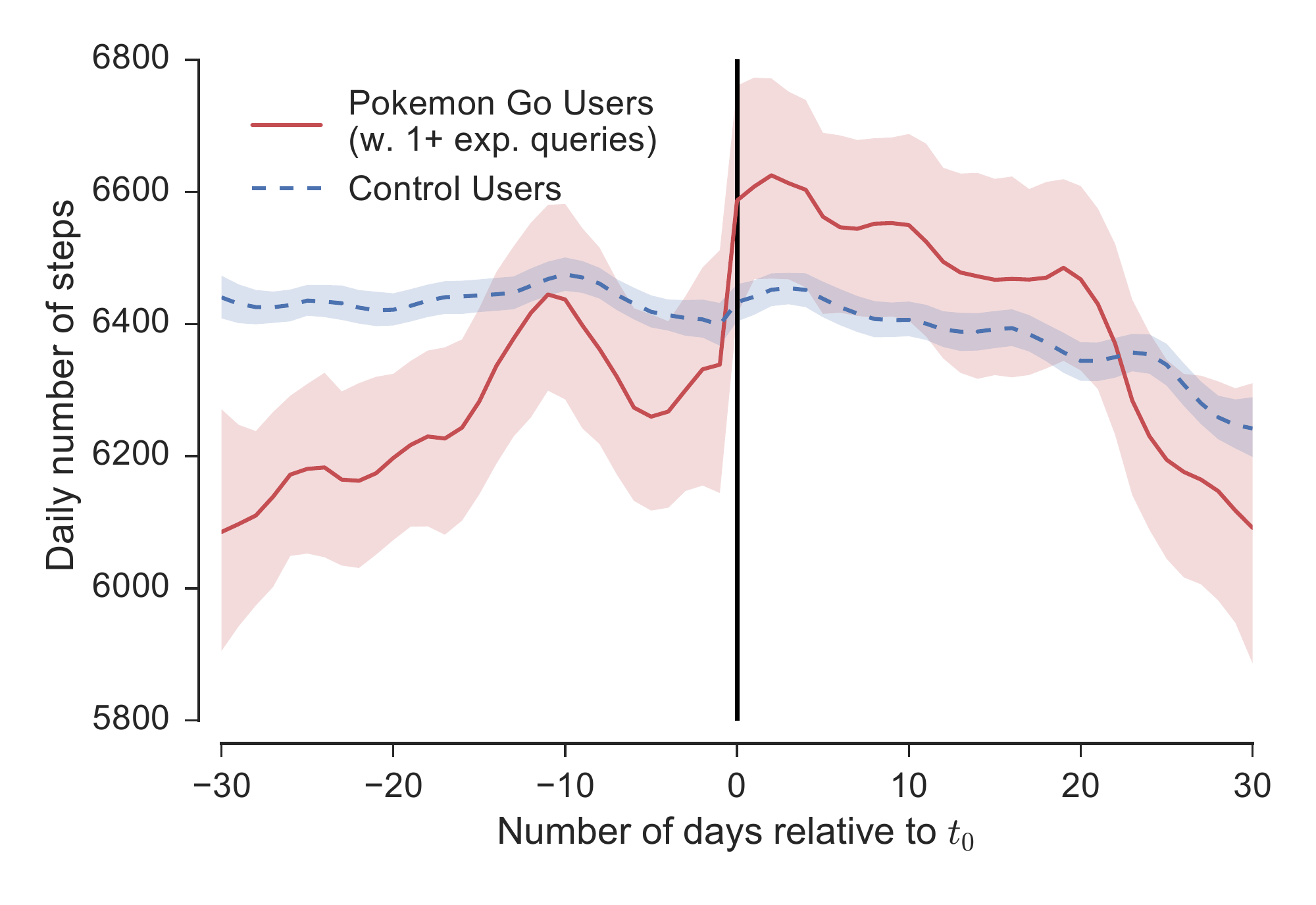}\\
\includegraphics[width=0.999\columnwidth]{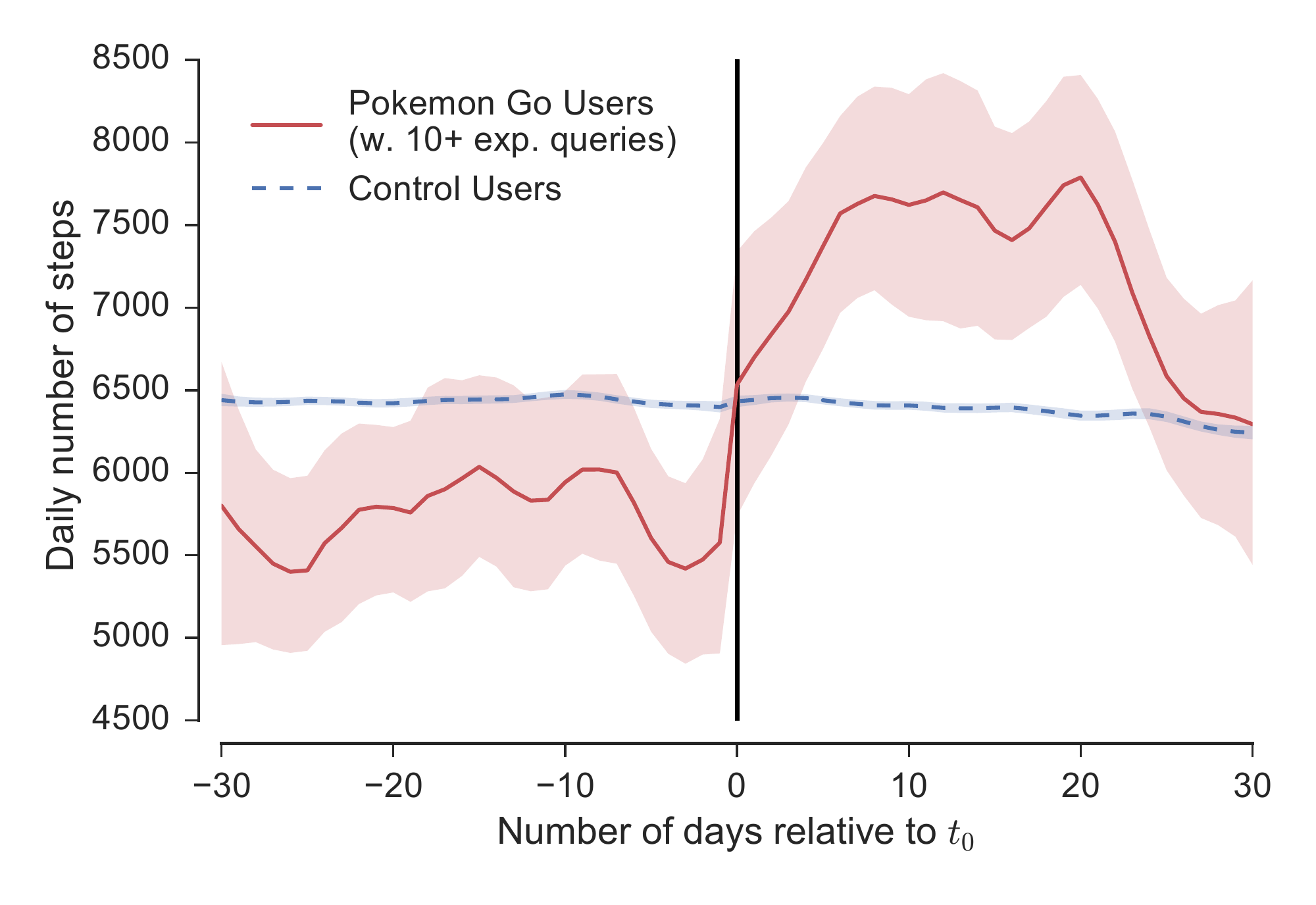}
\caption{Effect of \pokemongo on physical activity. 
Plots show daily steps in absolute numbers for both \pokemongo users (red) and control users (blue).
Top plot shows effect for users with at least one experiential query. 
Bottom plot shows effect for users with at least ten experiential queries.
In particular for the users who show significant interest in \pokemongo (bottom), we observe large average increases of 1473 steps or 26\% over the 30 days following on the first experiential query.
Over the same time, the control group (same for both plots) decreased their activity by 50 daily steps on average.
Error bars (shaded) in this and all following plots correspond to bootstrapped 95\% confidence intervals~\cite{efron1994introduction}.
}
\label{fig:group_effect_time_series}
\end{figure}

\section{Results}
\label{sec:effect_on_pa}

We now present results on the influence of \pokemongo usage on physical activity.
We study longitudinal physical activity data in Section~\ref{subsec:longitudinal_analysis}.  
We quantify the dose-response relationship between interest in \pokemongo and physical activity in Section~\ref{subsec:quantifying_effect_size}. 
Next, we examine potentially heterogeneous treatment effects by examining various subgroups based on several demographic attributes in Section~\ref{subsec:effect_size_user_demographics}.
We compare \pokemongo to four popular mobile health apps in terms of their effect on physical activity in Section~\ref{subsec:app_comparison}.
Lastly, we quantify the impact of \pokemongo on public health in Section~\ref{subsec:public_health_impact}.

\subsection{Longitudinal Analysis}
\label{subsec:longitudinal_analysis}

Starting to play \pokemongo is associated with significant increases in physical activity.
Changes in average activity level over time are illustrated in Figure~\ref{fig:group_effect_time_series}.
The top plot shows activity for \pokemongo users with at least one experiential query and the bottom plot shows activity for \pokemongo users with at least ten experiential queries (\ie, users who expressed significant interest in details of \pokemongo commands and operation).

We observe a significant increase in physical activity after the first experiential query for \pokemongo users compared to the control group.
The control group slightly decreased their activity by 50 daily steps on average ($p<10^{-20}$; we use Mann--Whitney U-Tests for hypothesis tests unless noted otherwise). 
In contrast, \pokemongo users increased their activity by 192 daily steps ($p<10^{-7}$).The plot shows a steep increase on the day of the first experiential query ($t_0$) suggesting that the observed increased activity indeed stems from engaging with \pokemongonospace.
%
We find that \pokemongo users initially have less activity than the average Microsoft Band user in the US (dashed blue line; 178 daily steps less; $p<10^{-20}$).
However, following the start of \pokemongo play, their activity increases to a level larger than the control group (65 daily steps more; $p<10^{-20}$).

The bottom row in Figure~\ref{fig:group_effect_time_series} shows similar but much larger effects for \pokemongo users with at least ten experiential queries; that is, users who showed significant interest in \pokemongonospace. 
These users are initially significantly less active than the average Microsoft Band user in the US, getting 5,756 daily steps compared to 6,435 daily steps in the control group ($p<10^{-20}$).
After they start playing \pokemongo they exhibit a large increase in activity to an average of 7,229 daily steps (1,473 daily steps difference; $p<10^{-15}$), which now is about 13\% larger than the control population ($p<10^{-20}$). 
This observation suggests that there is a dose-response relationship between interest in \pokemongo and the effect on physical activity, which we analyze in detail in Section~\ref{subsec:quantifying_effect_size}.

We note that increases in steps before $t_0$ could stem from starts with the game in advance of queries about \pokemongonospace, as we are using the first experiential query as a proxy for the start of play.  If users begin to play without ever issuing a search query about \pokemongonospace, we could see increases in activity before $t_0$.
However, since we observe steep increases in activity exactly at $t_0$, this suggests that the proxy for starting is valid for most users.

Note that physical activity for both \pokemongo user groups (top and bottom row) decreases again after about three to four weeks after the first experiential query. 
However, also note that the activity for the more strongly engaged group (bottom) drops down to a higher level than they started out with. 
This suggests that there could be a longer-term behavior change and that future work is needed to study long-term effects of \pokemongonospace.

\subsection{Dose-Response Relationship between \pokemongo and Physical Activity}
\label{subsec:quantifying_effect_size}

\begin{figure}[t]
\centering
\includegraphics[width=1.0\columnwidth]{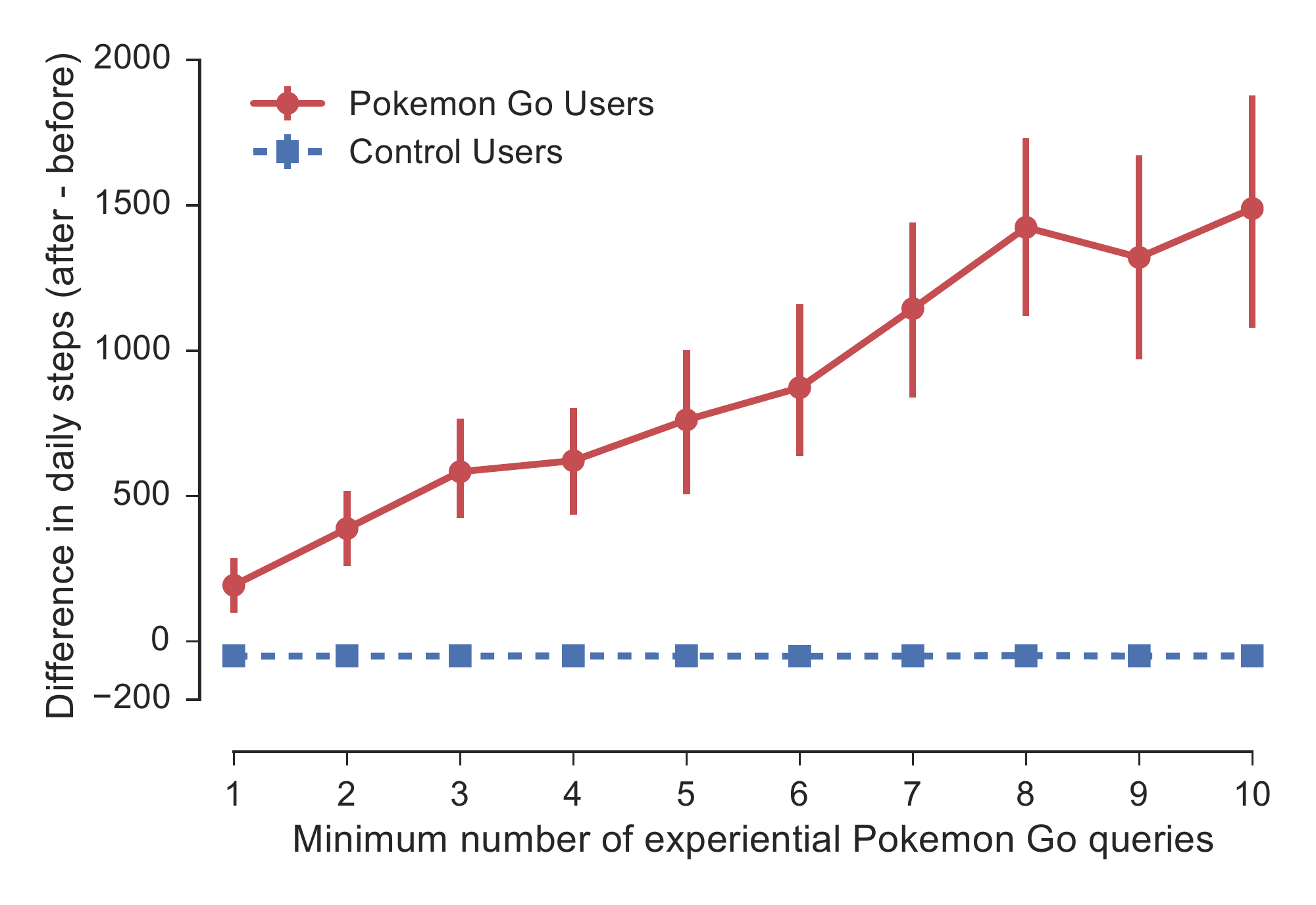}
\caption{Effect sizes measuring the difference in average number of daily steps between the periods before and after $t_0$, for different user populations based on the minimum number of experiential queries; that is, towards the right the group of users becomes smaller and more and more interested in \pokemongonospace.
At any level there are significant differences between the effect for \pokemongo users (red) and the control users (blue).
The effect increases linearly with the number of \pokemongo queries. 
This dose-response relationship between expressed interest in \pokemongo and physical activity suggests that these users are in fact playing \pokemongo and that playing the game makes them more active.
Confidence intervals for control group are too small to be visible.
}
\label{fig:effect_size_summary}
\end{figure}

We find that users that are more engaged with \pokemongo exhibit larger increases in physical activity (see Figure~\ref{fig:effect_size_summary}).
For users that expressed any interest in \pokemongo we find significant increases in activity compared to the control group which decreases their activity by 50 steps a day. 
Further, we find that these increases in steps scale roughly linearly with the number of experiential queries from 192 daily steps increase (3\%) for users with one or more experiential queries up to an increase of 1473 daily steps (26\%) for users with ten or more experiential queries.


Furthermore, the linear increase in physical activity with the number of experiential \pokemongo queries strongly suggests that activity increases observed in users querying a search engine for \pokemongo are causally explained by their engagement with \pokemongonospace.
If there were other confounding factors that explained the difference in activity between our \pokemongo group and the control group over time and those changes had nothing to do with \pokemongonospace, then one would not expect to find such a clear dose-response relationship as given in Figure~\ref{fig:effect_size_summary}.

\begin{figure*}[ht]
\centering
\includegraphics[width=0.90\textwidth]{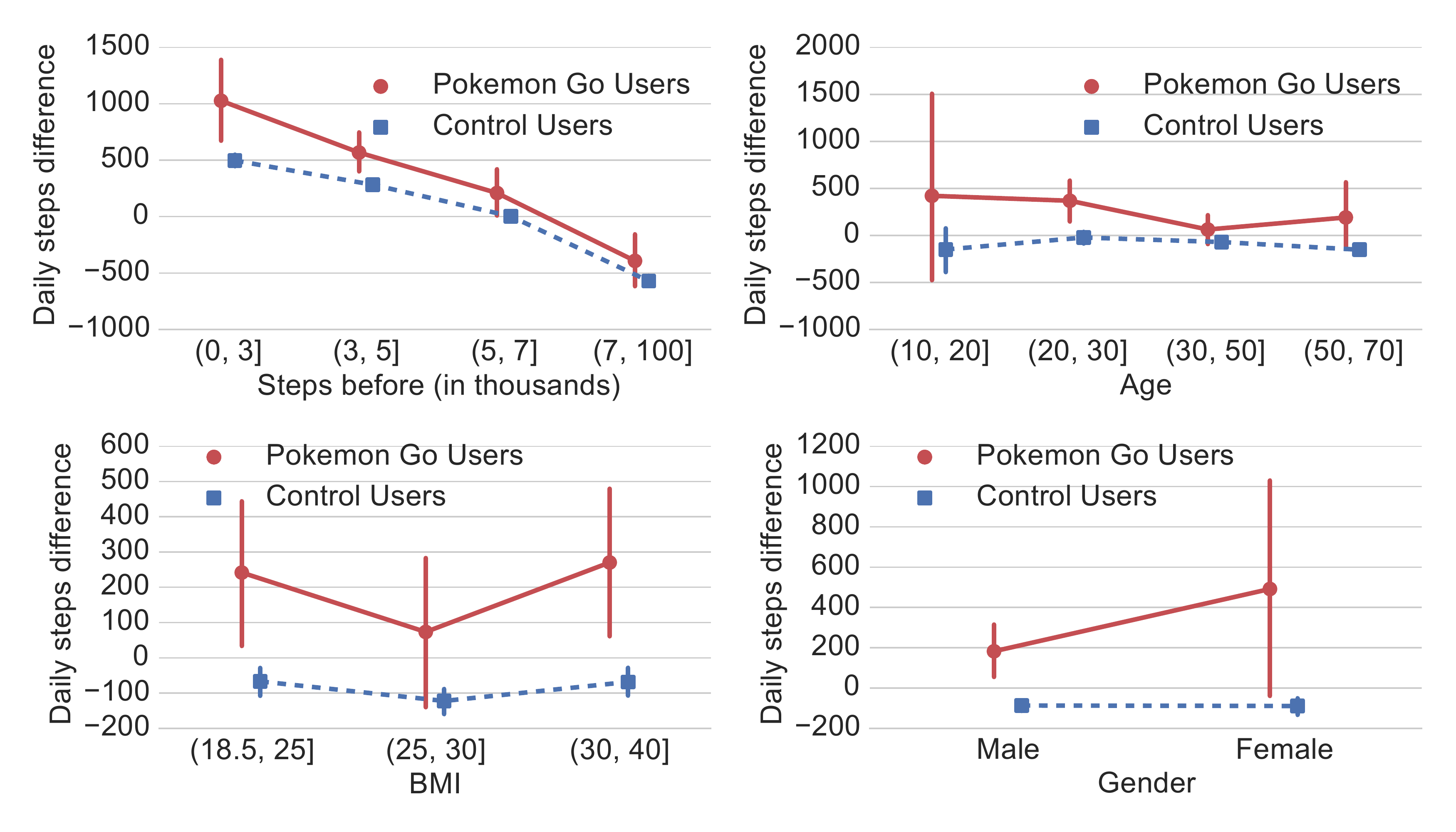}
\caption{
Effect sizes of physical activity increase or decrease by user demographics, including prior physical activity level (top left), age (top right), body mass index (BMI; bottom left), and gender (bottom right). 
In all cases, we find that \pokemongo users (red) exhibit larger changes then their respective control group (blue; see Section~\ref{subsec:effect_size_user_demographics}).
These results suggest that physical activity increases due to \pokemongo are not restricted to particular subgroups of users but widely spread across the overall study population.
}
\label{fig:effect_size_user_demographics}
\end{figure*}

\begin{figure}[ht!]
\centering
\includegraphics[width=1.0\columnwidth]{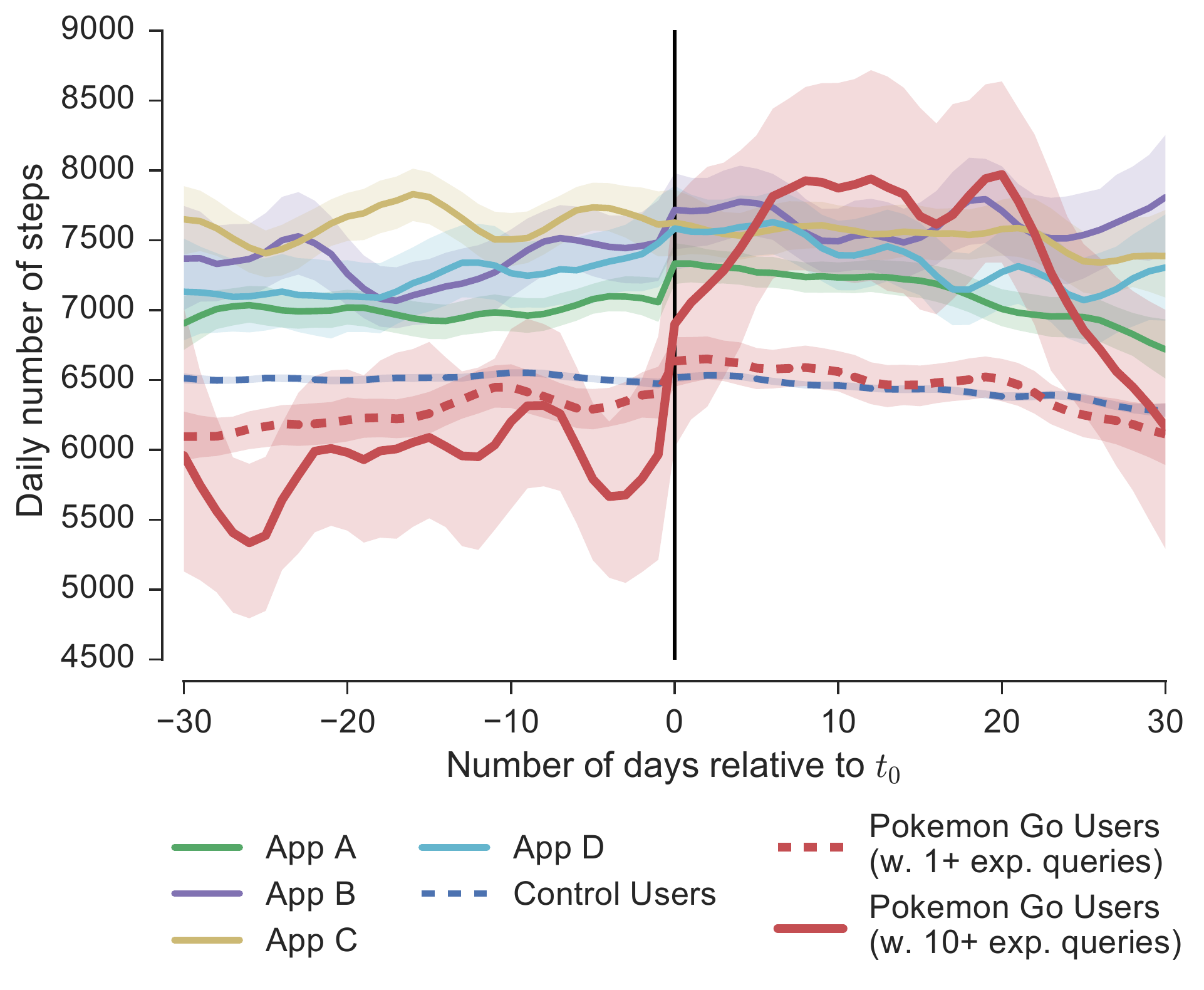}
\caption{Comparing the effect of \pokemongo app to leading consumer health apps (A, B, C, D).
\pokemongo users are less active than the average wearable user (Control) before starting to play but see larger increases in physical activity compared to the four consumer health apps.
}
\label{fig:app_comparison}
\end{figure}

\textbf{}

\subsection{Does Everyone Benefit?}
\label{subsec:effect_size_user_demographics}

Since this analysis is on user level, we only consider users who track their activity at least seven days before and after $t_0$.
Overall, the \pokemongo users \textit{increased} their activity by 194 daily steps ($p<0.01$; Wilcoxon Signed-Rank-Test).
Over the same time period, the control users \textit{decreased} their activity by 104 steps ($p<10^{-20}$; Wilcoxon Signed-Rank-Test).
%
Figure~\ref{fig:effect_size_user_demographics} illustrates the effect size split by previous activity level (top left), age (top right), body mass index (BMI; bottom left) and gender (bottom right).

We find that \pokemongo has increased physical activity across men and women of all ages, weight status, and prior activity levels.
%
In particular, we find that both \pokemongo users and control users who are very inactive exhibit large activity increases and users who are relatively active even exhibit a decrease in activity on average.
However, we find that \pokemongo users exhibit larger effects than the control across all levels of prior activity (all $p<0.025$).
We find the largest differences between the two groups for users that previously were sedentary (\ie, below 5,000 daily steps~\cite{tudor2004many}).
%
Furthermore, \pokemongo users exhibit bigger increases in activity than control users across all age groups (all $p<0.040$; except 10-20 year old group which was small) though we find largest effects for younger users between 10 and 30 years.
%
We also find that the positive effect on physical activity does not vary much across all BMI groups, which is encouraging since obese individuals (30 < BMI $\leq$ 40) are typically less active than healthy subjects~\cite{dishman1996increasing}.
The activity differences in the \pokemongo groups were always larger than the differences in the control group across all BMI groups (all $p<0.021$). 
%
Lastly, we find that activity differences in the \pokemongo groups were larger than the differences in the control group for both men and women (all $p<0.022$).
Increases for women were not significantly different from increases for men ($p=0.110$; note small sample size for women).

In summary, we find that \pokemongo increased activity all across the studied population, largely independent of prior activity level, age, weight status, or gender. 
These results are encouraging since they suggest that any positive effects due to \pokemongo are available even to sedentary, obese,  and older users.  
Effectively reaching these users with physical activity interventions is critical for public health~\cite{AmericanHeartAssociationNews2016}. 

\subsection{Comparison to Existing Health Apps}
\label{subsec:app_comparison}

\pokemongo leads to larger increases in physical activity than other mobile health apps and further attracts more users who are not yet very active.
The average daily steps over time is visualized in Figure~\ref{fig:app_comparison} (using same smoothing method as before). 
First, we observe that users of all four health apps are significantly more active than the average wearable user (6,514 daily steps) even before starting to use the health app (6,997-7,616 daily steps; see activity before $t_0$ in Figure~\ref{fig:app_comparison} for apps A,B,C,D).
\pokemongo users were less active than the average user (5,901-6,265 daily steps).
This demonstrates that \pokemongo app is attracting a different group of users which is less active and therefore would see larger health benefits from improving their activity~\cite{cdc2014pafacts,us2008physical}.
The temporal pattern for the health apps do not contribute strong evidence that these apps are leading to significant behavior change. 
One exception is app A with its users significantly increasing their activity at day 0. 
However, this increase in activity is lower compared to the effect of \pokemongonospace.
Users of app A increased their activity on average by 111 daily steps or 1.6\%.
Compare this to 194-1502 daily steps or 3.1-25.5\% for \pokemongo users with at least one or ten experiential queries, respectively.
In particular, users demonstrating large engagement with \pokemongo exhibit much larger increases in activity than users of any other app in our comparison.

These results emphasize the special contribution that activity-encouraging games could have on physical activity and public health.
These games attract a wide range of people including those with low prior physical activity.
We have demonstrated throughout this paper that such games can lead to significant activity increases.

\subsection{Estimating the Public Health Impact of \newline\pokemongonospace}
\label{subsec:public_health_impact}

\xhdr{Effect on US Physical Activity}
On average, users with an experiential query for \pokemongo increased their physical activity by 192 steps a day for the next 30 days (see Section~\ref{subsec:quantifying_effect_size}).
Extrapolating this average effect size to 25 million \pokemongo users in the US~\cite{surveymonkey2016ususers}, we find that \pokemongo added 144 billion steps within the first 30 days to US physical activity.
This is equivalent to walking around the equator 2,724 times or 143 round trips to the moon.

\xhdr{Effect on Meeting Activity Guidelines}
Using all users tracking steps at least seven days before and after their first experiential query for \pokemongonospace, 
we find that that the fraction of users meeting physical activity guidelines (\ie, getting 8,000 average daily steps~\cite{tudor2011stepsguidelines, tudor2011accelerometer}) stays approximately constant for users with one or more experiential queries (22.0\% before vs. 21.9\% after $t_0$) and control users (24.1\% before vs. 23.5\%).
However, for highly engaged \pokemongo users with at least ten experiential queries, we find that during the 30 days after they start playing 160\% more users achieve 8,000 average daily steps (12.2\% before vs. 31.7\% after; relative increase of 160\%).
For comparison, 21\% of US adults meet these guidelines~\cite{cdc2014pafacts,us2008physical}.

\xhdr{Effect on Life Expectancy}
We found that more engaged users exhibited average physical activity increases of up to 1,473 daily steps (see Section~\ref{subsec:quantifying_effect_size}). 
This substantial impact on exercise across the society could have a measurable impact on US life expectancy due to well-established health benefits of physical activity~\cite{ding2016economic,lee2012effect,miles2007physical,sparling2000promoting,WHO2010parecommendation}.
If we assume that \pokemongo users, between 15 and 49 years old, would be able to sustain an activity increase of 1,000 daily steps,
this would be associated with 41.4 days of additional life expectancy.
Across the 25 million US \pokemongo users~\cite{surveymonkey2016ususers}, this would translate to 2.825 million years additional life added to US users.

\section{Discussion}
\label{sec:discussion}

The \pokemongo phenomenon has reached millions of people overnight and dominated news media for weeks after its release~\cite{surveymonkey2016ususers,TechCrunch2016PokemonUserStats,guardian2016worldwideusers}. 
Health professionals have pointed out potential benefits including increased physical activity, spending more time outside and exploring the neighborhood and city, social interactions, and mastering game challenges but have also raised concerns such as injury, abduction, trespassing, violence, and cost~\cite{AmericanHeartAssociationNews2016,serino2016pokemon}.
In this study, we have precisely quantified the impact of \pokemongo on physical activity and studied the effect on different groups of individuals.

\subsection{Principal Results}
We find that playing the game significantly increased physical activity on the group-level (see Section~\ref{subsec:longitudinal_analysis} and Section~\ref{subsec:quantifying_effect_size}) as well as the individual-level (see Section~\ref{subsec:effect_size_user_demographics}) over an observation period of approximately four weeks.
The more interest the users showed in \pokemongo (measured through intensity of search queries seeking details about game usage), the larger the increase of physical activity (see Section~\ref{subsec:quantifying_effect_size}). 
For example, users that issued ten \pokemongo queries on details of the game within the two months after release of the game, increased their activity by 1479 steps a day or 26\%.
These increases are not restricted to already active and healthy individuals but also reach individuals with low prior activity levels and overweight or obese individuals.
Comparing \pokemongo to existing mobile health apps, we find further evidence that \pokemongo is able to reach low activity populations while mobile health and fitness apps largely draw from an already active population (see Section~\ref{subsec:app_comparison}).
This highlights the promise of game-based interventions versus traditional approaches, which have often been ineffective for these groups of people~\cite{dishman1985determinants,marshall2004challenges}.

Given its great popularity, \pokemongo has significantly impacted US physical activity and added an estimated 144 billion steps to US physical activity which is about 2,724 times around the world or 143 round trips to the moon.
Furthemore, highly engaged users were almost three times as likely to meet official activity guidelines in the 30 days after starting to play \pokemongo compared to before.
If this user engagement could be sustained, \pokemongo would have the potential to measurably affect US life expectancy.

Our study shows the large potential impact that activity-en\-cour\-aging games could have on society. 
However, we have also highlighted challenges in realizing this potential.
Most importantly, games would need to be able to sustain long-term engagement and lead to sustained behavior change. 
Furthermore, these games might not be appealing to everyone (\eg, we observed males to be more likely to play the games than females), and clearly these games should not replace but complement existing physical activity programs (\eg,~\cite{dobbins2009school,marshall2004challenges,reis2016scaling,sallis1998environmental,salmon2007promoting,sparling2000promoting}).
Understanding how to design games and how to bring together games and health interventions will be important to public health in the future.
As a first step, our study helps to provide guidance on what could come of continuous engagement and with additional engagement.

\subsection{Limitations}
Out study is not without limitations. 
First, the study population is not a random sample of US population. Subjects were able to afford a wearable device for activity tracking and willing to share their data for research purposes. 
Further, we use individuals search queries as a proxy for playing \pokemongo and consider the number of queries as indicating the degree of engagement. However, we find strong evidence that the proxies for usage and engagement are effective.
The method identifies a fraction of users that is very similar to to independent estimates of \pokemongo penetration in the US (see Section~\ref{subsec:identifying_users}) and we find a strong dose-response relationship between the number of \pokemongo queries and increased physical activity (see Section~\ref{subsec:quantifying_effect_size}).
Lastly, our follow-up period is currently restricted to 30 days. 
Future work is needed to study the long-term effectiveness of games such as \pokemongo to increase physical activity.

\subsection{Comparison with Prior Work}
The link between physical activity and improved health outcomes has been well-established (\eg,~\cite{ding2016economic,lee2012effect,miles2007physical,sparling2000promoting,WHO2010parecommendation}).  
At the same time, only a small fraction of people in developed countries meet official physical activity guidelines~\cite{cdc2014pafacts,us2008physical}.
Consumer wearable devices for activity tracking are becoming more prevalent in the general population. The devices can enable us to better understand real-world physical activity and how to best support and encourage healthier behaviors~\cite{hayden2016mobile,servick2015mind}.

Few research studies to date have harnessed data obtained from consumer wearables to study influences of the devices on physical activity. 
However, a number of medical studies have examined accelerometer-defined activity (\eg,~\cite{troiano2008physical,tudor2004many}), rather than relying on self-report measures. 
Studies have found that use of pedometers and activity trackers for self-monitoring can help increase activity~\cite{thorup2016cardiac,wang2016mobile} but other studies have reported mixed results~\cite{wang2015wearable}.
Beyond enabling self-monitoring, encouraging additional activity through reminders lead to increased activity only for the first week after the intervention and did not lead to any significant changes after six weeks in a randomized controlled trial~\cite{wang2015wearable}.

To encourage healthy behavior change, researchers have studied the design of ``exergames''~\cite{gobel2010serious,sinclair2007considerations,staiano2011exergames}, video games combined with exercise activity, and location-based games where game play progresses through a player's location~\cite{avouris2012review}.
However, no such game has been nearly as popular and widely used as \pokemongonospace.
%
Such games have yet to be integrated into physical activity programs, even though one US college recently announced a physical education class based on \pokemongo~\cite{ryssdal2016pokemoncollegeclass}.

There is a growing body of work on using large-scale search query logs to identify subjects with particular conditions for research studies, including such efforts as detecting adverse reactions to medications and identifying signals that could help with screening for cancer~\cite{paparrizos2016screening,white2014health,white2016early}.
Other work has studied activity-related posts on social media to better understand the sharing of health behaviors~\cite{kendall2011descriptive,park2016persistent,teodoro2013fitter} but has not yet connected such data to ground-truth health behaviors or focused on interventions on a large scale.

To the best of our knowledge, this is first study of the link between the usage of \pokemongo or similar games on physical activity and health. Also, this is the first effort to combine data from wearable devices with information drawn from search engine queries.

\subsection{Conclusions}
Novel mobile games which require players to physically move in the real world appear to be an effective complement to traditional physical activity interventions and they are able to reach millions of engaged users. 
We studied the effect of \pokemongo on physical activity through a combination of large-scale wearable sensor data with search engine logs, 
and showed that the game leads to significant increases in physical activity over a period of 30 days, with particularly engaged users increasing their average activity by 1,473 steps a day or 26\%.  Based on our findings, we estimate that the game has already added an estimated 144 billion steps to US physical activity.  If engagement with \pokemongo could be sustained over the lifetime of its many users, we estimate that the game would add an estimated 2.825 million years of additional lifetime to its US users.
We see great promise for public health in designing geocentric games like \pokemongo and in working to sustain users' engagement with them.

\xhdr{Acknowledgments}
The authors thank Jen Hicks for feedback on the manuscript.



\balance
\bibliographystyle{abbrv}
\bibliography{refs}


\end{document}